# Improving the Time Stability of Superconducting Planar Resonators

M.S. Moeed,[1,2] C.T. Earnest,[1,2] J.H. Béjanin,[1,2] A.S. Sharafeldin,[1,2] and M. Mariantoni[1,2]

[1]*Institute for Quantum Computing, University of Waterloo, 200 University Avenue West, Waterloo, Ontario N2L 3G1, Canada*

[2]*Department of Physics and Astronomy, University of Waterloo, 200 University Avenue West, Waterloo, Ontario N2L 3G1, Canada*

## ABSTRACT

*Quantum computers are close to become a practical technology. Solid-state implementations based, for example, on superconducting devices strongly rely on the quality of the constituent materials. In this work, we fabricate and characterize superconducting planar resonators in the microwave range, made from aluminum films on silicon substrates. We study two samples, one of which is unprocessed and the other cleaned with a hydrofluoric acid bath and by heating at 880 ℃ in high vacuum. We verify the efficacy of the cleaning treatment by means of scanning transmission electron microscope imaging of samples' cross sections. From 3 h-long resonator measurements at $\approx 10\,\text{mK}$ and with $\approx 10$ photonic excitations, we estimate the frequency flicker noise level using the Allan deviation and find an approximately tenfold noise reduction between the two samples; the cleaned sample shows a flicker noise power coefficient for the fractional frequency of $\approx 0.23 \times 10^{-15}$. Our preliminary results follow the generalized tunneling model for two-level state defects in amorphous dielectric materials and show that suitable cleaning treatments can help the operation of superconducting quantum computers.*

## INTRODUCTION

Quantum computers hold the promise of solving classically intractable algorithms such as the factorization of large integers, quantum search and optimization, and quantum simulations [1]. Among other systems, superconducting devices as well as trapped ions and semiconductor devices are the leading candidates for the experimental implementation of a practical quantum computer [2]. Superconducting devices [3], in particular, are already being used in medium-scale quantum computers comprised of ~100 quantum bits (qubits).

Superconducting planar devices are implemented by patterning micro- and nano-metric aluminum (Al) structures on silicon (Si) substrates (wafers), with fabrication methods similar to those used by the semiconductor industry. These devices are operated at a temperature $T \approx 10$ mK and are used to store the electromagnetic energy of fields oscillating at a frequency $f_0 \sim 5$ GHz.

Large-scale superconducting quantum computers will be operated continuously for extended time periods ranging from hours to days [4]. The time stability of the qubit parameters is, thus, a critical requirement to realize practical quantum algorithms. In recent studies [5-8], it has been shown that the energy $E_0 = hf_0$ and energy relaxation rate $\Gamma$ of superconducting qubits suffer from significant time fluctuations that may hinder the implementation of quantum algorithms. The main process responsible for such fluctuations is likely due to the interaction between the qubit and *two-level state* (TLS) defects [9], which are hosted in amorphous dielectric materials such as Si and Al oxide. These materials surround the qubit circuitry because the surface of the Si substrate and the top surface of the Al structures are covered with a few nanometers of native Si and Al oxide, respectively.

There exists a large body of theoretical and experimental work on the interaction between TLS defects and superconducting devices (for an in-depth review see, e.g., Ref. [9]), where both superconducting qubits and resonators have been studied. Planar resonators can be implemented by fabricating a lumped-element capacitor $C$ and inductor $L$ on a Si substrate and by connecting them in parallel, thus forming a harmonic oscillator with resonance frequency $f_0 = 1/(2\pi\sqrt{LC})$; similarly, a coplanar waveguide (CPW) transmission line of finite length $\ell$ and open at both ends serves as a distributed-element resonator with fundamental resonance frequency $f_0 = \bar{c}/(2\ell)$, where $\bar{c}$ is the velocity of light in the Si-vacuum medium (see, e.g., chapter 7 in Ref. [10]). Planar qubits are realized by connecting resonators to Josephson tunnel junctions, which act as nonlinear inductors, thus forming anharmonic oscillators [3].

Temperature $T$ and excitation power $P$ strongly affect the behavior of the TLS defects. In the case of a resonator, for example, the total resonator dielectric loss is given by [11]

$$\delta = \frac{1}{Q_\text{i}} = \delta_\text{TLS}^0 \frac{\tanh\frac{hf_0}{2k_\text{B}T}}{\left(1 + \frac{\langle n_\text{ph}\rangle}{\langle n_\text{ph}\rangle^\text{c}}\right)^\alpha} + \delta^* \quad , \tag{1}$$

where $Q_\text{i}$ is the intrinsic (or internal) quality factor of the resonator, $\delta_\text{TLS}^0$ the TLS loss at $T = 0$ K and $\langle n_\text{ph}\rangle = 0$, $k_\text{B}$ the Boltzmann constant, $\langle n_\text{ph}\rangle$ the mean photon number circulating in the resonator, $\langle n_\text{ph}\rangle^\text{c}$ a critical mean photon number above which the TLS defects start to be saturated, $\alpha \sim 0.5$, and $\delta^*$ accounts for all non-TLS losses. For high-quality resonators $Q_\text{i} \rightarrow +\infty$ and $P \sim hf_0 \langle n_\text{ph}\rangle \kappa_\text{c}$, where $\kappa_\text{c} = 2\pi f_0/Q_\text{c}^*$ and $Q_\text{c}^*$ is the rescaled coupling quality factor of the resonator (see, e.g., chapter 7 in Ref. [10]). The resonator is operated in the *quantum regime* when $k_\text{B}T \ll hf_0$ and $\langle n_\text{ph}\rangle \sim 0$, in which case $\delta \cong \delta_\text{TLS}^0 + \delta^*$. The total dielectric loss $\delta$ can be obtained by measuring the transmission scattering coefficient $S_{21}$ of the resonator, which is a Lorentzian function at frequencies $f \sim f_0$; the Lorentzian amplitude is a peak for half-wave resonators (open-open CPW transmission lines) and a dip for quarter-wave resonators (open-short CPW transmission lines). Following Ref. [11], $Q_\text{i}$, $Q_\text{c}^*$, and $f_0$ are obtained by fitting the complex function

$$\frac{1}{S_{21}} = 1 + \frac{Q_i}{Q_c^*} e^{i\phi} \frac{f_0}{f_0 + i2Q_i(f - f_0)} \quad , \tag{2}$$

where $\phi$ is an offset angle (also fitted) and $i^2 = -1$.

In our work of Ref. [11] (similar to the works in Refs. [12-15]), we have fabricated and characterized *loss* in CPW resonators made from thin-film Al on Si substrates. In particular, we have cleaned a Si substrate before Al deposition using a chemical treatment based on a hydrofluoric (HF) acid followed by a physical treatment based on a thermal annealing at 880 °C in high vacuum. Comparing similar resonators on the sample with substrate cleaning to one without cleaning, we have showed a $Q_i$ improvement by $\approx 256\%$ in the quantum regime. In this article, we consider the same samples and characterize the cleaning treatment with scanning transmission electron microscope (STEM) imaging of samples' cross sections. We study the samples' *noise* properties at $T \approx 10$ mK and with $\langle n_{\text{ph}} \rangle \approx 10$, i.e., the dependence with time $t$ of $f_0$ and $Q_i$. We analyze the time series $f_0(t)$ and $Q_i(t)$ and, thus, $\Gamma(t)$ by means of standard metrology tools such as the Allan deviation (ADEV) and the power spectral density (PSD) [16].

Comparing the unprocessed and cleaned samples, we find an approximately tenfold reduction in the frequency flicker $(1/f)$ noise accompanied by an approximately twofold reduction in the time-averaged energy relaxation rate $\langle \Gamma \rangle$. These findings corroborate the model in Ref. [17], although our annealing temperature is about three times higher than in that work and is performed prior to metal deposition and our measurements taken at a slightly higher $\langle n_{\text{ph}} \rangle$.

## EXPERIMENTS

### Sample preparation

We fabricate two samples on high-resistivity ($> 10$ k$\Omega$ cm) 500 μm-thick undoped Si substrates (for more details on sample preparation see our work of Ref. [11]):
1. One control sample without any Si substrate surface cleaning ("Unprocessed");
2. One sample with two sequential Si substrate surface cleanings ("RCA 1 + HF + 880 °C Anneal"):
   a. A chemical cleaning consisting of an RCA Standard Clean-1 (RCA 1) process followed by a 1 % HF dip for 1 min, ~10 min prior to step b;
   b. A physical cleaning consisting of a thermal annealing step at $880 \mp 15$ °C for 10 min in high vacuum (HV) at a pressure $\lesssim 4 \times 10^{-7}$ mbar prior to Al deposition.

Aluminum films with 100 nm thickness are deposited on the Si substrates in an ultra-HV (UHV) electron-beam physical vapor-deposition system from Plassys-Bestek SAS, model MEB 550 SL3-UHV, at a deposition pressure of $\sim 2 \times 10^{-8}$ mbar.

The Al films are patterned with optical lithography and etched with an inductively coupled plasma (ICP) etcher from Oxford Instruments plc, model Plasmalab System 100 ICP 380 III-V and metals chlorine etcher. After etching, any remaining chlorinated species on the Al and Si surfaces are removed by soaking them in water for $\gtrsim 10$ min. The remaining photoresist is removed with Remover PG, a proprietary N-methyl-2-pyrrolidinone-based stripper from the MicroChem Corp. Each substrate is finally diced forming square samples with a lateral dimension of 15 mm.

Figure 1 shows optical microscopy images of a sample similar to those studied in this work. The samples have an equal layout, where one feed CPW transmission line is capacitively coupled to ten meandered quarter-wave resonators; each resonator has a different $f_0$ in a multiplexed design (see the work of Ref. [11] for more details). We measure $S_{21}$ through the feed line in the frequency range between 4 and 8 GHz. For comparison, we select one resonator from each sample designed with the same $f_0 = 4.5$ GHz; the measured time-averaged resonance frequencies $\langle f_0 \rangle$ (see Table II) are slightly different due to fabrication imperfections. Each resonator is capacitively coupled to the corresponding feed line with a designed coupling strength $\kappa_c = 40$ kHz. The feed line and resonator feature a center conductor of width $S = 15$ μm and gaps of width $W = 9$ μm.

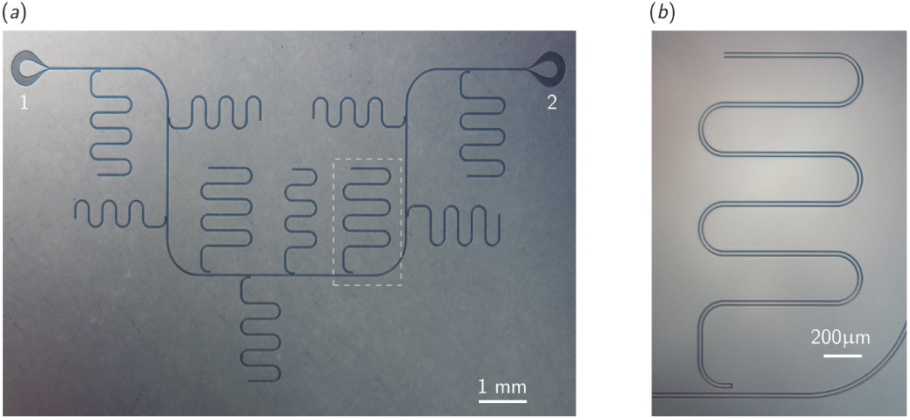

Figure 1. Optical microscopy images of a sample with the same layout and main fabrication steps as those studied in this work. (a) Ten meandered CPW resonators coupled to one feed line in a multiplexed design; each resonator has a different $f_0$. The resonator designed with $f_0 = 4.5$ GHz studied here is enclosed by a white dashed box. The input and output ports, 1 and 2, used to measure $S_{21}$ are indicated below the corresponding pads, which mate with the Pogo pins of a quantum socket [18]. (b) Detail of the resonator in the white dashed box in (a), showing the capacitive coupling to the feed line (bottom region) as well as the shorting to ground of the resonator centered conductor (top region; end of resonator). Scale markers are indicated in both panels.

### Resonator measurements

Each sample is housed in a quantum socket package as described in the work of Ref. [18]. The package is thermalized to the mixing chamber of a dilution refrigerator (DR) with a base temperature $T \approx 10$ mK and is enclosed within magnetic and infrared shields. Input and output microwave coaxial cables connect the package to the room-temperature electronics. The total attenuation of the input line is $\approx 76$ dB at 5 GHz. The output line includes two cascaded circulators thermalized to the mixing chamber stage and one circulator to the still stage ($T \approx 800$ mK) of the DR followed by an amplifier thermalized to the 3 K stage of the DR and one room-temperature amplifier; the cold amplifier is characterized by a nominal gain of $\approx 39$ dB and a noise temperature of $\approx 5$ K in the frequency range of interest. All measurements are performed with a vector network analyzer (VNA) from Keysight Technologies Inc., model PNA-X N5242A.

We measure all $S_{21}$ traces in the frequency range centered around 4.5 GHz with a span of ~50 kHz and with $\langle n_{\rm ph} \rangle \approx 10$ for both the Unprocessed and RCA 1 + HF + 880 °C Anneal samples; each trace is comprised of 417 points. The VNA intermediate frequency (IF) bandwidth $\Delta f_{\rm IF}$ is reported in Table I for each sample. The total measurement time of one $S_{21}$ trace, $\Delta t$, is given by the observation time $\tau_{obs} \sim 417/\Delta f_{IF}$ and the dead time $\tau_d$ before and after each observation, $\Delta t = \tau_{obs} + \tau_d$. The dead time before an observation is due to measurement triggering and after to data downloading; additionally, the dead time after includes $S_{21}$-trace fitting. Both $\tau_{obs}$ and $\tau_d$ slightly vary from observation to observation and so does $\Delta t$. Each $S_{21}$ trace is measured once.

Table I. Measurement parameters for both samples (see main text for details). $\Delta f_{IF}$: VNA's IF bandwidth; $N$: Number of measured $S_{21}$ traces; $t_{tot}$: Total measurement time of all traces; $\langle \Delta t \rangle$: Mean value of all $\Delta t$s, with margins of error (one standard deviation) in parenthesis; $\tau_0$: Maximum measurement time of one $S_{21}$ trace in the series; $\langle \tau_{obs} \rangle$: Mean value of all $\tau_{obs}$s; $\langle CI(f_0) \rangle$: Mean value of the CIs for a 68% confidence level of all $f_0$s; $\langle Q_i \rangle$: Time-averaged $Q_i$; $\sigma_{Q_i}$: Standard deviation of $Q_i$; $\langle CI(Q_i) \rangle$: Mean value of the CIs for a 68% confidence level of all $Q_i$s; $\langle Q_c^* \rangle$: Time-averaged $Q_c^*$. All numerical entries are rounded down, except for $\Delta f_{IF}$ and $N$. Note that the $\tau_0$ entry for the RCA 1 + HF + 880 °C Anneal sample is significantly larger than the $\langle \Delta t \rangle$ error margin because it accounts for the removal of one outlier (see main text) between two consecutive series' elements; all other entries are reported without accounting for outlier removal.

| Sample | $\Delta f_{IF}$ (Hz) | $N$ — | $t_{tot}$ (s) | $\langle \Delta t \rangle$ (s) | $\tau_0$ (s) | $\langle \tau_{obs} \rangle$ (s) | $\langle CI(f_0) \rangle$ (Hz) | $\langle Q_i \rangle$ $10^6$ | $\sigma_{Q_i}$ $10^6$ | $\langle CI(Q_i) \rangle$ $10^3$ | $\langle Q_c^* \rangle$ $10^5$ |
|---|---|---|---|---|---|---|---|---|---|---|---|
| Unprocessed | 200 | 4377 | 10797.24 | 2.47(1) | 2.59 | 1.86 | ∓100 | 0.59 | 0.07 | ∓20 | 3.49 |
| RCA 1 + HF + 880 °C Anneal | 100 | 2530 | 10797.82 | 4.27(5) | 8.74 | 3.70 | ∓82 | 1.32 | 0.16 | ∓74 | 3.37 |

We acquire $S_{21}$-trace measurements consecutively forming an unevenly time-spaced series with $N$ elements for a total measurement time of all traces, $t_{tot} = \sum_{j=0}^{N-1} \Delta t(j)$, with $j \in \mathbb{N}$ and $\Delta t(0) = 0$ s ($t_{tot} \approx 3$ h). We fit each element of the series using Equation (2) obtaining the time series $\{t_j, f_0\}$ and $\{t_j, Q_i\}$, where $t = t_j = \sum_{j \leq (N-1)} \Delta t(j)$, and $f_0$ and $Q_i$ are fitting parameters. For each sample, $N$, $t_{tot}$, the mean value of all values of $\Delta t$ in the series, $\langle \Delta t \rangle$, with margins of error, as well as the maximum measurement time of one $S_{21}$ trace in the series, $\Delta t_{max} \equiv \tau_0$, are reported in Table I. Additionally, we report the mean value of all values of $\tau_{obs}$ in the series, $\langle \tau_{obs} \rangle$, as well as the time-averaged $Q_i$ and $Q_c^*$, $\langle Q_i \rangle$ and $\langle Q_c^* \rangle$, the standard deviation of $Q_i$, $\sigma_{Q_i}$, and the mean value of the confidence intervals (CIs) of all values of $f_0$ and $Q_i$ in the series, $\langle CI(f_0) \rangle$ and $\langle CI(Q_i) \rangle$, respectively. The frequency bandwidth of our measurement system is $f_h \sim 1/\langle \Delta t \rangle \sim 10^{-1}$ Hz; despite being narrow, such a bandwidth is sufficient to study slow-varying TLS defect dynamics that result, e.g., in flicker noise. The time series $\Delta f_0(t) = f_0(t) - \langle f_0 \rangle$ is plotted in Figure 2 (a) for the Unprocessed and RCA 1 + HF + 880 °C Anneal samples. We compute $\Gamma(t) = 2\pi f_0(t)/Q_i(t)$, which is plotted in Figure 2 (b) for both samples.

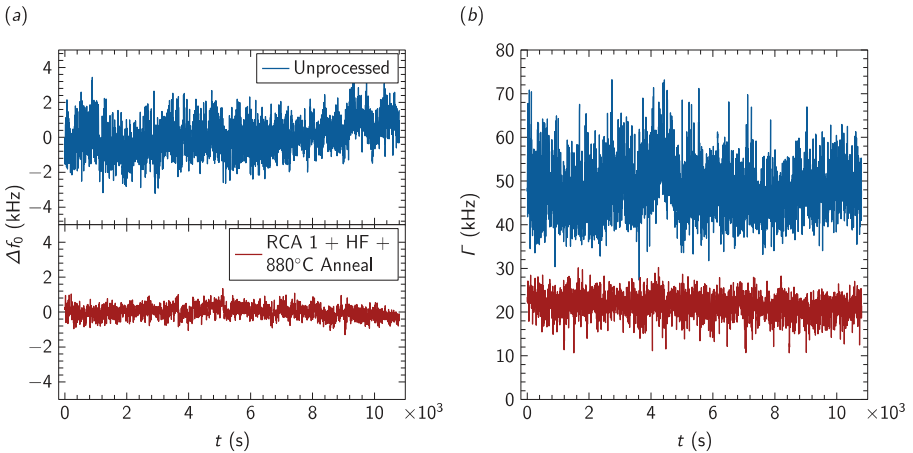

Figure 2. Time series for both samples. (a) $\Delta f_0$ vs. $t$; the time series are displayed on separate axes. (b) $\Gamma$ vs. $t$.

By inspecting the time series in Figure 2 (a) and (b), we do not observe any significant systematic (i.e., deterministic) effects on the data such as linear drifts over time. Thus, we can proceed with the statistical analysis, or noise characterization, of the signals.

## RESULTS AND DISCUSSION

### Sample characterization

The Al films are deposited on the Si substrates immediately after each cleaning treatment, thus providing an excellent protecting layer from any interactions with the environment. Hence, the most suitable region to characterize the effects of each substrate cleaning treatment is the substrate-metal (SM) interface. We study this interface by means of cross-sectional STEM imaging.

Samples' cross sections or lamellae are prepared using a focused ion beam (FIB) system from Thermo Fisher Scientific Inc., model Helios G4 plasma FIB (PFIB) UXe DualBeam. The PFIB system works on xenon+ (Xe+) ions, which are ideal to study interfaces; Xe+ ions, in fact, do not implant at the interfaces, thus avoiding the contamination resulting from traditional gallium (Ga)-based FIBs [19]. The samples' surfaces are protected by depositing a mixture of carbon and platinum (Pt) prior to PFIB milling. For each sample, two trenches are milled on both sides of the area of interest using a Xe+ beam. The resulting lamellae are lifted out from the samples using a micromanipulator needle. The lamellae are then attached to a copper TEM grid using Pt. Finally, the lamellae are thinned to a thickness of ~80 nm and treated with a low-voltage cleaning at 2 kV.

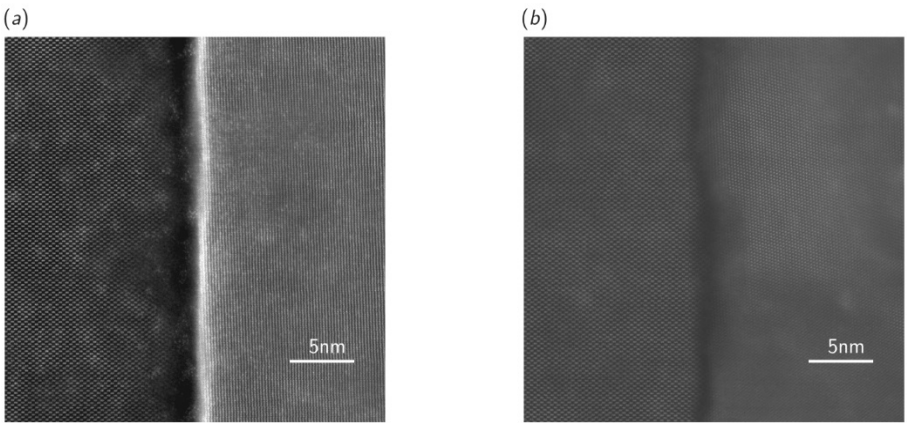

Figure 3. High-resolution STEM images of the SM interface lamellae. For both samples, Si is on the left-hand side and Al on the right-hand side of the interface. (a) Unprocessed sample. (b) RCA 1 + HF + 880 °C Anneal sample. A clear reduction of any interfacial contaminants and, thus, TLS defects is visible between the samples. Scale markers are indicated in both panels.

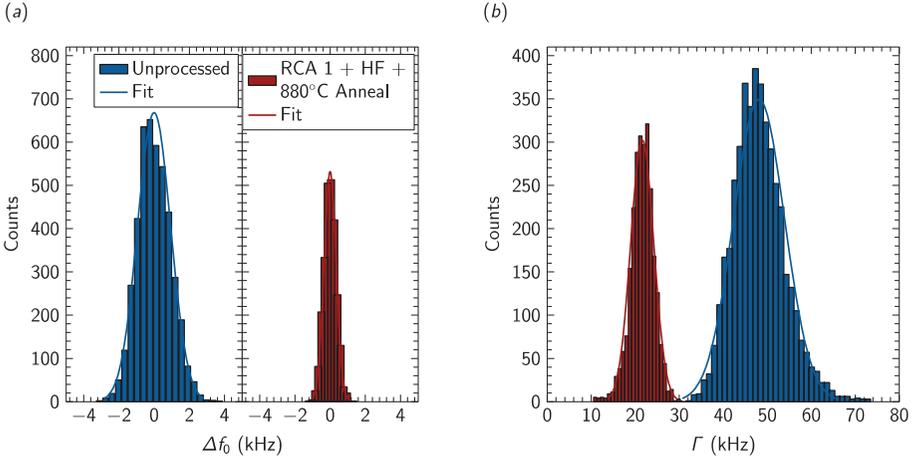

Figure 4. Histograms (bins) with normal distribution fitting curves (solid lines) of the time series for both samples (see Table II). (a) Counts vs. $\Delta f_0$; the histograms, which are displayed on separate axes, show a spread reduction between the two samples. (b) Counts vs. $\Gamma$; the histograms show both a location and a spread reduction between the two samples.

The lamellae are imaged in a TEM system from the FEI Company, model Titan 80-300 HB. The system, which is operated at 200 kV, is equipped with an image and probe corrector from CEOS GmbH and a Gatan imaging filter from Gatan, Inc. (type "Quantum Energy Filter"). We acquire high-angle annular dark-field high-resolution STEM images of the SM interface, which are shown in Figure 3 for both samples. The Unprocessed sample shows an ~2 nm-thick interfacial layer likely hosting TLS defects. The RCA 1 + HF + 880 °C Anneal sample, instead, shows a clean interface; this interface is rougher than for the Unprocessed sample due to the heating process. In our previous work of Ref. [11], we have shown the electron energy loss spectroscopy maps of a small region at the SM interface for both samples, using, however, the less ideal Ga+ FIB to prepare the lamella for the RCA 1 + HF + 880 °C Anneal sample. Additionally, in that work we have studied other interfaces such as the substrate-air (vacuum) interface and the effects of roughness.

### ADEV estimation and noise model

Figure 4 shows $f_0(t)$ and $\Gamma(t)$ as histograms for both the Unprocessed and RCA 1 + HF + 880 °C Anneal samples. The time-averaged resonance frequencies $\langle f_0 \rangle$ and energy relaxation rates $\langle \Gamma \rangle$ as well as the corresponding standard deviations $\sigma_{f_0}$ and $\sigma_\Gamma$ of each histogram are reported in Table II. All histograms fit well to a normal distribution.

A powerful tool to perform the statistical analysis of a random signal is the ADEV [16] [20]. The ADEV can be plotted as a function of the observation time $\tau$. A simple inspection of the ADEV slopes makes it possible to readily identify the main noise processes associated with the experiment and, thus, to perform a quantitative analysis by fitting to the corresponding noise models [21].

Typically, the time series acquired in experiments are comprised of equally spaced times. As explained in Subsection Resonator measurements we acquire unevenly spaced time series, from which we generate equally spaced series following the procedure outlined below. We note that the time spacing in our series *is not* highly irregular (see Table I); hence, our procedure to estimate the ADEV remains largely unbiased.

We first remove any outliers from the time series by discarding values that are five times larger than the median absolute deviation of all values in the series. The outliers correspond to $S_{21}$ traces that fail to fit to Equation (2). We remove 8 elements from $f_0(t)$ and 20 elements from $\Gamma(t)$ for the RCA 1 + HF + 880 °C Anneal sample and no elements from either series for the Unprocessed sample.

We then compute for each sample the fractional $f_0(t)$ and $\Gamma(t)$ offsets from the corresponding $\langle f_0 \rangle$ and $\langle \Gamma \rangle$ [16] [20], resulting in the fractional time series

$$y(t) = \frac{f_0(t) - \langle f_0 \rangle}{\langle f_0 \rangle} \quad \text{and} \quad z(t) = \frac{\Gamma(t) - \langle \Gamma \rangle}{\langle \Gamma \rangle}$$

respectively. We then process the data as follows:
1. We define an observation time bin ("$\tau$ bin") as $\tau \equiv k\tau_0$, with $k \in \mathbb{N}_{>0}$;
2. For each value of $\tau$, we truncate a given fractional time series such that the total measurement time of the truncated series, $\tilde{t}_{\text{tot}}$, is divisible by one $\tau$ bin. The series must be truncated by removing the least number of elements from its end. The total number of $\tau$ bins within the series is thus $M(\tau) = \tilde{t}_{\text{tot}}/\tau$;
3. We choose values of $\tau$ such that there exist at least two $\tau$ bins within a given truncated fractional time series;
4. We split up the truncated fractional time series into an integer number $M$ of $\tau$ bins for each value of $\tau$;
5. For each $\tau$ bin, we average the $y$ or $z$ values of the series falling within that bin. For the $y(t)$ series, for example, we obtain the approximate time-averaged fractional frequency time series

$$\bar{y}_m(\tau = k\tau_0) \cong \frac{1}{L_m} \sum_{\ell=1}^{L_m} (y_m)_\ell \quad , \tag{3}$$

where $m \in \mathbb{N}_{>0}$ is the index spanning the set of $\tau$ bins within the series, $L$ is the number of series' elements within the $m$th bin, $\ell \in \mathbb{N}_{>0}$, and $y$ is one value in the subset of $y$ values that fall within the $m$th bin and $(y\ )_\ell$ the $\ell$th value in that subset;
6. Equation (3) finally allows us to compute an estimator of the Allan variance, which for $y(t)$ reads [16]

$$\sigma_y^2(\tau) \cong \frac{1}{2(M-1)} \sum_{m=1}^{M-1} [\bar{y}_{m+1}(\tau) - \bar{y}_m(\tau)]^2 \quad . \tag{4}$$

7. The square root of Equation (4) is the estimated ADEV for $y(t)$, $\sigma_y(\tau)$. Similarly, we obtain the estimated ADEV for $z(t)$, $\sigma_z(\tau)$.

The estimated ADEVs $\sigma_y(\tau)$ and $\sigma_z(\tau)$ are shown in Figure 5 (a) and (c) for both samples.

In addition to the ADEV, we estimate the PSD associated with the time series for both samples. We use the Welch's method with 50% overlap and a Hamming window (see, e.g., chapter 14 in Ref. [22]), obtaining the estimated PSDs $S_y(f)$ and $S_z(f)$ for $y(t)$ and $z(t)$, respectively, which are displayed in Figure 5 (b) and (d).

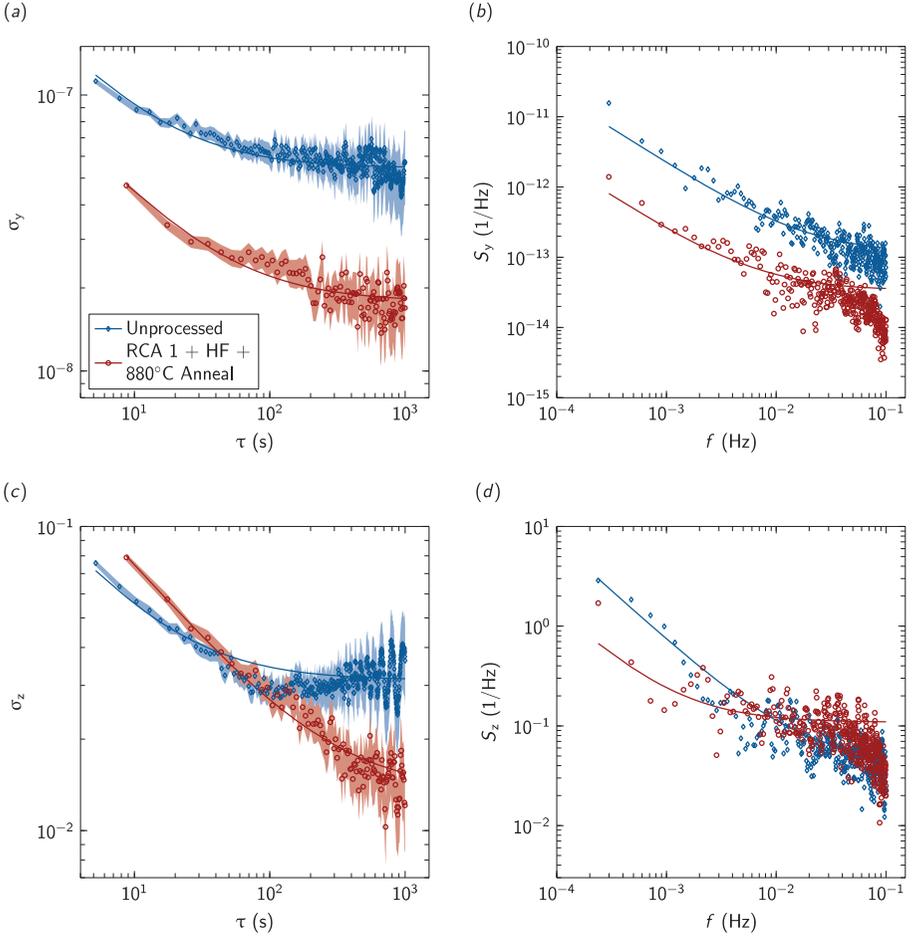

Figure 5. Estimated ADEVs and PSDs (symbols) with fitting curves (solid lines) for both samples (see main text for details). The margins of error (standard error) associated with the ADEV values are indicated by shaded regions. (a) $\sigma_y$ vs. $\tau$. (b) $S_y$ vs. $f$. (c) $\sigma_z$ vs. $\tau$. (d) $S_z$ vs. $f$. Note that for the ADEVs the white noise limit (slope = $-0.5$) is on the left-hand side of the plot, whereas the flicker noise limit (slope = 0) on the right-hand side. For the PSDs the white noise limit (slope = 0) is on the right-hand side of the plot, whereas the flicker noise limit (slope = $-1$) on the left-hand side. The flicker noise level is significantly reduced in all cases, whereas only the frequency white noise level is reduced (see Table II).

In order to quantify the results of Figure 5 (a) and (c), we fit the Allan variances for both $y(t)$ and $z(t)$ to a noise model comprised of white and ideal flicker noise [16]

$$\sigma_{y,z}^2(\tau) = \frac{h_0^{y,z}}{2\tau} + h_{-1}^{y,z} 2 \ln 2 \quad , \qquad (5)$$

where $h_0^{y,z}$ and $h_{-1}^{y,z}$ are the power coefficients for the white and flicker noise, respectively; for all fits, we use the Levenberg-Marquardt algorithm. The numerical values of all fitted power coefficients along with their corresponding CIs for a 95% confidence level are reported in Table II for both samples. The fitting curves associated with the values in Table II are overlaid to the data in Figure 5 (a) and (c).

By means of the $\mu - \alpha$ mapping from $\tau$ to $f$ as defined in frequency metrology [16], it is possible to derive the equivalent noise model for the PSDs from Equation (5),

$$S_{y,z}(f) = h_0^{y,z} + \frac{h_{-1}^{y,z}}{f} \quad . \tag{6}$$

We use the same power coefficients fitted from Equation (5) to plot the curves associated with Equation (6), which are overlaid to the data in Figure 5 (b) and (d). Notably, these curves match the data well, confirming the validity of the model and fitting method. Conversely, an $\alpha - \mu$ mapping from $f$ to $\tau$ gives similar results (not shown).

One of the assumptions behind the noise model of Equations (5) and (6) is $\tau \gg 1/(2\pi f_h)$ [16]. This assumption is not met for the data close to the origin in Figure 5 (a) and (c). These data regions capture the white noise level in our experiments. Thus, $f_h$ may have an effect on the white noise estimates $h_0$ reported in Table II.

Table II. Noise parameters for both samples. For the first four entries, the values are obtained from the histograms in Figure 4. $\langle f_0 \rangle$: Time-averaged $f_0$; $\sigma_{f_0}$: Standard deviation of $f_0$; $\langle \Gamma \rangle$: Time-averaged $\Gamma$; $\sigma_\Gamma$: Standard deviation of $\Gamma$; $h_0^y$: Power coefficient of the white noise for $y(t)$; $h_{-1}^y$: Power coefficient of the flicker noise for $y(t)$; $h_0^z$: Power coefficient of the white noise for $z(t)$; $h_{-1}^z$: Power coefficient of the flicker noise for $z(t)$. For all power coefficients, the CIs for a 95% confidence level are reported in parenthesis.

| Sample | $\langle f_0 \rangle$ (GHz) | $\sigma_{f_0}$ (kHz) | $\langle \Gamma \rangle$ (kHz) | $\sigma_\Gamma$ (kHz) | $h_0^y$ $10^{-14}$ (s) | $h_{-1}^y$ $10^{-15}$ | $h_0^z$ $10^{-2}$ (s) | $h_{-1}^z$ $10^{-4}$ |
|---|---|---|---|---|---|---|---|---|
| Unprocessed | 4.49 | 0.94 | 48.13 | 5.80 | 10.27(05) | 2.15(04) | 4.70(24) | 6.96(22) |
| RCA 1 + HF + 880 °C Anneal | 4.51 | 0.34 | 21.59 | 2.62 | 3.37(25) | 0.23(01) | 10.84(25) | 1.34(12) |

## **Comparison to related work**

In recent literature [21] [7], a more advanced ADEV estimator has been adopted: The overlapping ADEV [20]. This estimator, for example, results in much smaller margins of error compared to those displayed in Figure 5 (a) and (c). While this is the preferred estimator for telecommunication standards, we prefer to avoid its usage as it may lead to post-processing misinterpretations. In fact, we generated random white and flicker noise in numerical simulations and computed the associated overlapping ADEV. The processed data should match exactly the noise model of Equation (5). Instead, the overlapping ADEV results in very distinct "hill-like" peaks that can only be explained by a Lorentzian noise process [7], which is however absent in the simulated data.

In the work of Ref. [15], a sample similar to our RCA 1 + HF + 880 °C Anneal sample is fabricated. Before presenting a comparison between our and their sample, it is worth pointing out some fabrication analogies as well as a few crucial differences. First, they perform the HF dip in a 2% HF solution, whereas we perform it in a 1% solution. Second, their thermal annealing step is carried out at only 300 °C while pumping an HV Plassys e-beam evaporator system, which reaches a base pressure of $1.1 \times 10^{-7}$ mbar when the Si substrate is cooled down to room temperature. Third, after depositing a 150 nm-thick Al film they oxidize the film's top surface exposing it to high-purity molecular oxygen (O) at 10 mbar for 10 min; this step passivates the Al surface before exposing it to the less pure atmospheric O. Last, they etch a 1 µm-deep trench in correspondence to the CPW Si gaps with 400 nm Al undercuts; our samples, instead, feature a modest ~100 nm-deep trench due to a slight Si overetching. In addition, they perform noise measurements for $f_0(t)$ only, but at an estimated $\langle n_{\text{ph}} \rangle \approx 1$. They find $h^y_{-1} \approx 3.5 \times 10^{-15}$, which is similar to our result for the Unprocessed sample. Our RCA 1 + HF + 880 °C Anneal sample, instead, is characterized by a fractional-frequency flicker noise level that is approximately one order of magnitude lower (see Table II). We attribute this difference partly to the different fabrication procedure and, most importantly, to the higher $\langle n_{\text{ph}} \rangle$ used to perform our measurements. Finally, we notice that our white noise level is significantly higher than their $h^y_0 \approx 2.5 \times 10^{-16}$; this finding is possibly affected by our narrow measurement bandwidth $f_h$.

The most promising result of our work is the approximately tenfold reduction in the fractional-frequency flicker noise level between the Unprocessed and RCA 1 + HF + 880 °C Anneal samples, $2.15/0.23 \approx 9.35$. The corresponding loss reduction between the two samples is, instead, only approximately a factor of two: When considering $1/\langle Q_i \rangle$, the reduction is $(1/0.59)/(1/1.32) \approx 2.24$; for $\langle \Gamma \rangle$, $48.13/21.59 \approx 2.23$. It is important to stress that the results in Table II are obtained with $\langle n_{\text{ph}} \rangle \approx 10$, whereas in the work of Ref. [17] with $\langle n_{\text{ph}} \rangle \approx 2$.

In the work of Ref. [17], similar findings are explained within the framework of the generalized tunneling model (GTM), which is an extension of the standard tunnelling model (STM). The STM was developed to describe the effect of an ensemble of TLS defects in glasses and amorphous dielectric materials [23-25]. In both models, a TLS defect is hypothesised to be a double-well potential with two metastable ground states separated by a barrier with energy $E_{\text{TLS}}$. The TLS defects are categorized into two classes: Coherent TLS (CTLS) defects, for which $E_{\text{TLS}} \gg k_B T$, and two-level fluctuator (TLF) defects, for which $E_{\text{TLS}} \ll k_B T$. The former are characterized by transition frequencies $\sim f_0$, whereas the latter have frequencies well below 1 GHz. In the STM, TLF defects weakly interact with CTLS defects, which, in turn, semi-resonantly interact with a resonator resulting in *loss $\Gamma$*. The GTM takes into account of strong dipole-dipole interactions between spatially close TLF and CTLS defects resulting in stochastic time fluctuations much larger than $\Gamma$ and, thus, *noise*. Under the reasonable assumption that our treatments uniformly clean the substrate's top surface from both species of TLS defects, a noise reduction larger than loss reduction is expected due to the strong interactions leading to noise.

In our work of Ref. [11], for the same two samples studied here we have found the exponent in Equation (1) to be $\langle \alpha \rangle \approx 0.31$, which is less than $\alpha = 0.5$ as predicted by the STM. This result is consistent with strongly interacting TLS defects [17] and, thus, further suggests the validity of the GTM for our samples. Additionally, we find an approximately fivefold reduction in the fractional–energy-relaxation flicker noise level between the two samples, $6.96/1.34 \approx 5.19$.

**CONCLUSIONS**

We fabricate and characterize superconducting CPW resonators with $f_0 \approx 4.5$ GHz made from Al thin films deposited on Si substrates. We study two samples: One Unprocessed sample, for which the Si substrate is not cleaned from any surface contaminant; one RCA 1 + HF + 880 °C Anneal sample, for which the Si substrate is cleaned chemically with an RCA 1 and an HF dip and then physically with a thermal annealing step at 880 °C in HV prior to Al deposition.

We take STEM images and measure one resonator on each sample at $T \approx 10$ mK and with $\langle n_{\text{ph}} \rangle \approx 10$ for $\approx 3$ h, obtaining the time series $f_0(t)$ and $Q_i(t)$. We estimate the fractional-frequency flicker noise level using the ADEV and find an approximately tenfold noise reduction between the two samples, with the RCA 1 + HF + 880 °C Anneal sample having a noise power coefficient of $\approx 0.23 \times 10^{-15}$.

In future work, we intend to reproduce our results by measuring multiple samples from different wafers treated with a chemical or a physical clean, or both, as well as multiple resonators with different $f_0$ on each sample. In particular, we intend to characterize in detail the effect of the thermal annealing step for a wide range of temperatures from $\sim 200$ to $900$ °C. We want to further investigate the connection between frequency and energy-relaxation noise to further corroborate the GTM. In addition, we want to perform all measurements with $\langle n_{\mathrm{ph}} \rangle \lesssim 1$ using a Pound-Drever-Hall–type locking technique [26] and with a much broader $f_h$. Finally, we want to repeat similar loss and noise measurements on superconducting qubits with tunable $f_0$.

## ACKNOWLEDGMENTS

This research was undertaken thanks in part to funding from the Canada First Research Excellence Fund (CFREF) as well as the Discovery and Research Tools and Instruments Grant Programs of the Natural Sciences and Engineering Research Council of Canada (NSERC). The devices were fabricated at the Quantum-Nano Fabrication and Characterization Facility at the University of Waterloo, Canada. TEM imaging was performed by Natalie Hamada and Andreas Korinek at the Canadian Centre for Electron Microscopy at McMaster University, Canada. We would like to acknowledge the Canadian Microelectronics Corporation (CMC) Microsystems for the provision of products and services that facilitated this research, including CAD tools and design methodology. The authors thank Jonathan Burnett for their fruitful discussions on frequency metrology.